# Tunable electron-electron interactions in LaAlO$_3$/SrTiO$_3$ nanostructures

Guanglei Cheng[1,4], Michelle Tomczyk[1,4], Alexandre B. Tacla[2], Hyungwoo Lee[3], Shicheng Lu[1,4], Josh P. Veazey[1][†], Mengchen Huang[1,4], Patrick Irvin[1,4], Sangwoo Ryu[3], Chang-Beom Eom[3], Andrew Daley[2], David Pekker[1,4], Jeremy Levy[1,4]*

[1]Department of Physics and Astronomy, University of Pittsburgh, Pittsburgh, PA 15260, USA

[2]Department of Physics and SUPA, University of Strathclyde, Glasgow G4 0NG, UK

[3]Department of Materials Science and Engineering, University of Wisconsin-Madison, Madison, WI 53706, USA

[4]Pittsburgh Quantum Institute, Pittsburgh, PA 15260, USA

[†] Current address: Department of Physics, Grand Valley State University, Allendale, MI 49401, USA

*jlevy@pitt.edu

**ABSTRACT**

**The interface between the two complex oxides LaAlO$_3$ and SrTiO$_3$ has remarkable properties that can be locally reconfigured between conducting and insulating states using a conductive atomic force microscope. Prior investigations of "sketched" quantum dot devices revealed a phase in which electrons form pairs, implying a strongly attractive electron-electron interaction. Here, we show that these devices with strong electron-electron interactions can exhibit a gate-tunable transition from a pair-tunneling regime to a single-electron (Andreev bound state) tunneling regime where the interactions become repulsive. The electron-electron interaction sign change is associated with a Lifshitz transition where the $d_{xz}$ and $d_{yz}$ bands start to become occupied. This electronically tunable electron-electron interaction, combined with the nanoscale reconfigurability of this system, provides an interesting starting point towards solid-state quantum simulation.**

I. INTRODUCTION

Quantum simulators – easily reconfigurable quantum many-body systems – have been proposed as an experimental tool for attacking a number of problems in physics and materials science ranging from mechanisms of high temperature superconductivity to the design of novel materials[1, 2]. Currently, ultracold atom systems are considered to be the strongest candidates for building a quantum simulator because they are microscopically well understood, and are highly controllable: (1) one can relatively easily reconfigure the potential for the atoms using optical lattices and (2) under appropriate conditions one can adjust atom-atom interactions using a Feshbach resonance[3]. A solid-state quantum simulator, in which one uses electrons in nanostructures as opposed to atoms in optical lattices, could potentially be advantageous for a range of applications, especially because the interaction energy scales are large enough that relevant temperature regimes can be reached with standard refrigeration, whereas the pico-eV energy scales in cold atoms make reaching such temperatures an ongoing experimental challenge. While tunable single-particle potentials have been demonstrated in a number of two-dimensional-electron-gas (2DEG) systems, for example electrostatically gated modulation doped GaAs/AlGaAs heterostructures[4] and nanotubes/semiconducting nanowires[5], adjusting the electron-electron interactions has proven to be much more difficult.

Complex oxide interfaces, where electrons interact very strongly, are a good system to look for tunable electron-electron interactions. A promising example is the strongly-correlated 2DEG at the $LaAlO_3/SrTiO_3$ (LAO/STO) interface[6]. This interface possesses a rich collection of properties including superconductivity[7, 8] and magnetism[9-11] that are indicative of attractive and repulsive interactions, respectively. When the thickness of LAO is reduced to 3 unit cells the interface becomes intrinsically insulating[12], but it can be locally switched between ON

(conducting) and OFF (insulating) states by "writing" and "erasing" with a voltage-biased atomic force microscope (c-AFM) tip[13]. Using these "write" and "erase" c-AFM procedures, a number of reconfigurable nanostructures can be created with extreme nanoscale precision (~2 nm)[14-18]. If electron-electron interactions can be tuned, then, in combination with arbitrary reconfigurability, this platform may offer the desired ingredients for realizing a solid state quantum simulator.

The complex electron-electron interactions at the LAO/STO interface are derived from the properties of the STO substrate. Doping bulk STO to a low carrier density ($10^{17}$ cm$^{-3}$) results in a superconductor with a small Fermi surface (Fermi temperature $T_F \sim 13$ K) and low superconducting critical temperature ($T_c < 0.3$ K)[19]. In a 1969 paper, Eagles argued that the superconductivity in low-density STO involves Bose-Einstein Condensation (BEC) of strongly paired electrons, in contrast to conventional Bardeen-Cooper-Schrieffer (BCS) superconductivity in which electron pairing is weak and electron pair size is much larger than the inter-electron spacing[20]. A direct consequence of the strong pairing theory is that above $T_c$ the electrons are no longer condensed but remain in tightly bound pairs. The general phenomenology of transitioning from strong to weak pairing interactions, known as the "BEC-BCS crossover", has been thoroughly investigated both theoretically and experimentally in ultracold atoms[20-23]. Recently, the hallmark of BEC-regime physics – electron pairing without superconductivity – was observed at the LAO/STO interface[24]. Specifically, it was found that electron pairs persist up to pairing temperatures of $T_p \sim 1-10$ K and magnetic fields of $B_p \sim 1-10$ T, far higher than the superconducting critical temperature $T_c \sim 0.3$ K and upper critical magnetic field $\mu_0 H_{c2} \sim 0.3$ T. The ratio of pairing temperature to Fermi temperature $T_p / T_F \sim 0.1-0.8$ is much larger than that of conventional BCS superconductors, indicating that the pairing interactions in low density STO are indeed quite strong and attractive, and hence are on the BEC side of the BEC-BCS crossover.

Here, we investigate electron-electron interactions at the LAO/STO interface by measuring transport through a quantum dot (QD) device fabricated by c-AFM lithography. Experiments utilize a superconducting single electron transistor (SSET) geometry, where the QD is proximity coupled to two superconducting nanowire leads and a side gate. This setup is geometrically similar to the one reported in Ref. [24], but here we investigate higher electron densities on the QD and different gap structures in the leads. We observe a dramatic change in the transport properties as we tune the electron density on the QD using electrostatic gating (by a sketched side gate). At low gate voltages (low electron densities on the QD) the transport occurs via strongly bound electron pairs, as previously reported in Ref. [24]. On the other hand, at high gate voltages (high electron densities on the QD) the transport changes to a conventional single-particle regime. The single-particle transport appears to be carried by conventional Andreev bound states (ABS) that are localized on the QD[25-27].

We ascribe the dramatic change in the transport properties through the QD to the change of an electron-electron interaction constant that depends on electron density. At low electron densities, the electron-electron interactions are strongly attractive. Low-energy excitations of the QD consist of adding or removing strongly bound electron pairs; hence, transport proceeds via resonant pair tunneling [Fig. 1(a), top panel]. At higher electron densities, the interactions become repulsive. In this regime, the low energy excitations of the QD consist of adding or removing a single electron from the dot [Fig. 1(a), bottom panel]. Coupling the QD to superconducting leads results in the formation of conventional ABS, which are responsible for electron transport in this regime.

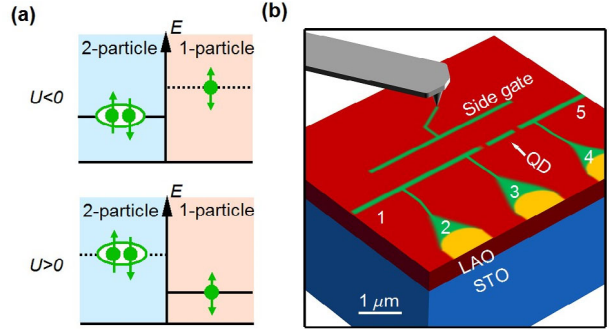

FIG. 1, **Superconducting single electron transistor (SSET).** (a) The excitation spectra is dependent on the interaction strength $U$ which is tunable by gate voltage. When $U<0$, the 2-electron ground state is lower than 1-electron ground state so that the QD favors adding a pair of electrons (top panel). Alternatively, when $U>0$, the QD favors adding a single electron (bottom panel). (b) Electron-electron interactions are probed by a SSET fabricated by c-AFM lithography. The nanowire QD is defined by two barriers between leads 3 and 4 separated by 1 µm. A side gate tunes the chemical potential of the QD.

## II. OBSERVATION OF PAIR AND SINGLE-PARTICLE TRANSPORT REGIMES

The SSET devices are fabricated by c-AFM lithography[24], as shown in Fig. 1(b). Using a voltage-biased c-AFM tip ($V_{tip} = 12$ V), we first "write" a nanowire network consisting of main channel leads (1 and 5) and three voltage sense leads (2, 3, and 4). The c-AFM tip is then directed to cut across the main channel with a small negative voltage applied ($V_{tip} = -0.3$ V) to engineer two tunnel barriers separated by 1 µm and located between leads 3 and 4. The tunnel barriers define the QD, and their strength determines the initial coupling strength to the leads. The nanowire section between leads 2 and 3 has no barriers and serves as a control wire. Finally, a side gate nanowire is written 1 µm away from the main channel to tune the chemical potential $\mu$, interaction strength $U$, and tunneling coefficient $t$. All of the nanowires have width $w\sim10$ nm at room temperature[13]. The entire setup can be regarded as a superconducting nanowire-QD-nanowire system.

Transport is measured in a four-terminal setup: we extract the differential conductance d$I$/d$V$ of the QD by passing a current through the main channel and simultaneously measuring the voltage drop between leads 3 and 4. Figure 2(a) shows the differential conductance d$I$/d$V$ of a typical SSET device as a function of the source-drain bias $V_{34}$ and side gate voltages $V_{sg}$ [see Fig. 1(b)] at low temperatures $T = 50$ mK and zero magnetic field ($B = 0$ T). Four distinct transport regimes can be identified in terms of $V_{sg}$ ranges: (i) well-defined conductance diamonds associated with resonant pair tunneling ($V_{sg} < -40$ mV), (ii) sub-gap transport via pair bound states ($-40$ mV $< V_{sg} < -30$ mV), (iii) sub-gap transport via Andreev bound states ($-30$ mV $< V_{sg} < -10$ mV) and (iv) Josephson transport ($V_{sg} > -10$ mV).

**(i) The well-defined conductance diamonds regime** ($V_{sg} < -40$ mV) is qualitatively similar to the transport reported in Ref. [24], in which we have associated the diamonds with resonant tunneling of strongly bound electron pairs. A series of zero-bias peaks (ZBP) are present near the "tips" of the diamonds as indicated in Fig 2(a). The ZBPs bifurcate as we increase the magnetic field above a critical value ($B_p \sim 1-2$ T), indicating the breaking of strongly bound pairs [Fig. 2(c)]. $B_p$ is typically much larger than the upper critical magnetic field $\mu_0 H_{c2} \sim 0.3$ T for destroying superconductivity[24].

The diamonds have a nearly insulating gap of roughly $4\Delta/e$, where $\Delta \sim 48$ μeV, in contrast to those observed in Ref. [24] without the insulating gap. Moreover, the diamonds are offset horizontally while still being connected by a straight line [see Fig 2(a)], which (as will be discussed below) indicates that the drain lead has gapless excitations while the source lead remains gapped. Such gapless excitations can arise from nanoscale imperfections (e.g., in carrier density), although the source and drain leads should be nominally identical. At sufficiently large magnetic fields, the pairing gap and the offset between the diamonds are simultaneously suppressed, see Fig. 2(b). The field (~1 T) at which the offset vanishes coincides with $B_p$ for electron pairing, suggesting the source lead is still gapped even when the superconductivity is suppressed above the upper critical field $\mu_0 H_{c2} \sim 0.3$ T.

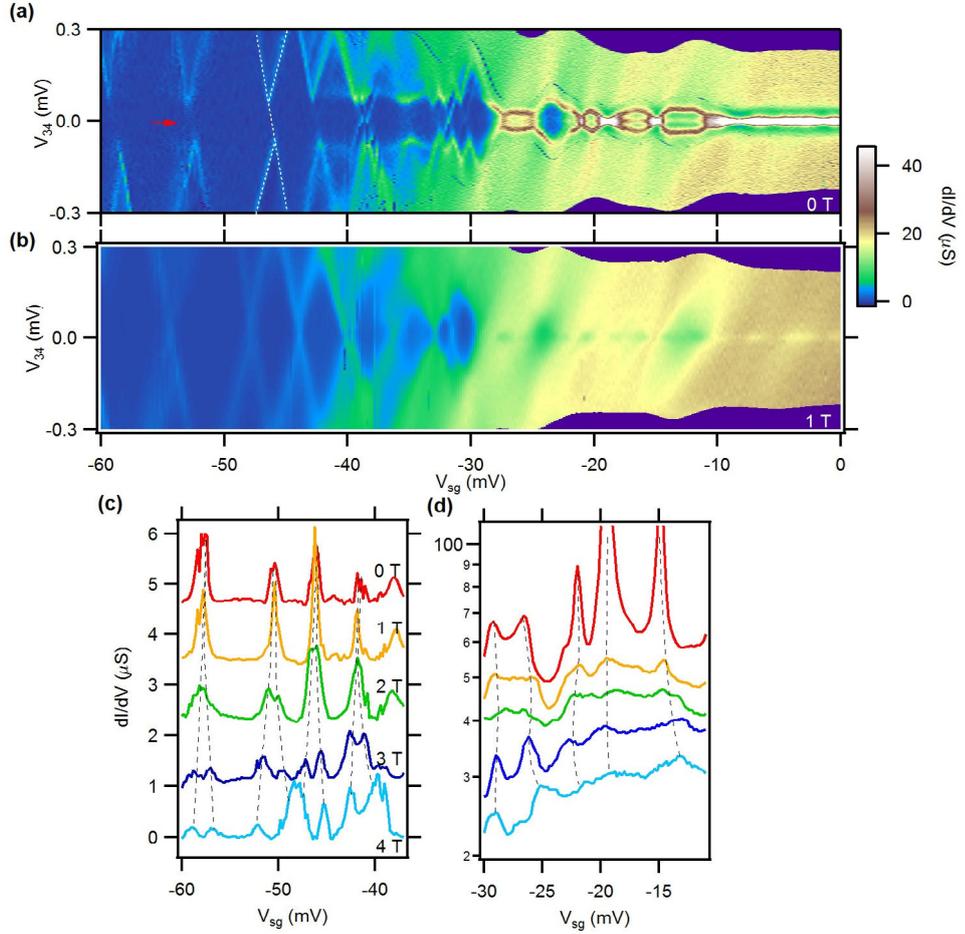

FIG. 2, Transport characteristics. At $T$=50 mK, d$I$/d$V$ is measured as function of $V_{34}$ and $V_{sg}$ at (a) $B$=0 T and (b) $B$=1 T. The dashed line in (a) is a guide to the eye showing how the diamonds are offset. The fact that the diamonds can be connected by a straight line indicates that one lead has a gap while the other is not gapped. The red arrow indicates the location of zero-bias peak. (c) Zero-bias line cuts at $B$=0 to 4 T in low $V_{sg}$ range (-60 mV<$V_{sg}$<-35 mV). The ZBPs bifurcate above $B_c$ (1~2 T), signifying pair tunneling. (d) Zero-bias line cuts at $B$=0 to 4 T in high $V_{sg}$ range (-30 mV<$V_{sg}$<-10 mV). The ZBPs do not bifurcate, signifying single electron tunneling.

**(ii) The sub-gap transport via pair bound states regime** ($-40$ mV $< V_{sg} < -30$ mV) is characterized by the appearance of relatively stronger conductance features inside the gap. These "X"-shaped features extend all the way across $4\Delta/e$ gap and appear to be particle-hole symmetric. We ascribe these features to pair bound states on the QD: electron pairs that are in a superposition of being a bound pair on the QD and in the superconducting lead.

**(iii) The sub-gap transport via ABS regime** ($-30$ mV $< V_{sg} < -10$ mV) is characterized by a dramatic change of the transport characteristics. The gap shrinks from $4\Delta/e$ to $2\Delta/e$ and at the same time the sub-gap features become much "brighter" ($dI/dV$ increases ~10-fold) as well as changing shapes from characteristic "X" features to "loop" features. We ascribe the dramatic change of the transport to the appearance of Andreev reflections. The absence of features at $V_{34} = 2\Delta/ne$, ($n$=1,3,4...) suggests that multiple Andreev reflection processes are irrelevant. Rather, the well-defined smooth loop features are a clear manifestation of transport via ABS.

In the diamond regime and the pair-bound state regime, the lowest excited state of the QD corresponds to adding (or removing, depending on $V_{sg}$) a pair of electrons from the dot. The emergence of ABS loops indicates the lowest excited QD level is characterized by adding (or removing) a single electron to the dot, as illustrated in Fig. 1(a). This assignment of the QD excitation structure can be further confirmed by examining the field dependence of the ZBPs. As shown in Fig. 2(d), no signs of ZBP bifurcation are observed up to $B = 4$ T in the ABS regime. In contrast, in the diamond regime the ZBPs bifurcate above $B_p \sim 1-2$ T. Since $B_p$ is generally decreasing with increasing $V_{sg}$ [24], this observation supports the conclusion that the origin of the ZBPs is single-particle in nature.

All the over 50 SSET devices we fabricated show electron pairing without superconductivity in the diamond regime. However, in order to observe closed ABS loops the QD has to be coupled to one gapped superconducting lead and one gapless "probe" lead. Although we did not purposefully design the gap structure in our devices, about 10% of the devices did have pronounced ABS loops. The existence of nanoscale imperfections which will sometimes make a particular lead gapless, is probably the primary factor in creating conditions necessary to observe ABS.

**(iv) The Josephson regime** ($V_{sg} > -10$ mV) appears at high side gate voltages (and hence, electron densities). In this regime the electron tunneling matrix element between the QD and the superconducting leads becomes large enough to enable coherent Josephson transport through the QD. The *I-V* characteristics in this regime are consistent with the RCSJ model[28, 29] of transport through a shunted Josephson junction with a typical critical current $I_c \sim 2.8$ nA (see Appendix IV).

## III. THEORETICAL MODEL OF TRANSPORT IN THE SSET

The experimental signatures of attractive and repulsive electron-electron interactions in transport can be well described by a minimal model of the SSET device. The ingredients for the model are (1) a superconducting lead with gapped excitations–which acts as a source of electron pairs; (2) a QD with a single-electron level of either attractive or repulsive interactions; (3) and a normal lead with gapless excitations–which acts as a sensor of electronic states on the QD. The reason for including both a gapless and a superconducting lead in the model is the fact that sketched LAO/STO nanowires tend to show at the same time both electron pairing and gapless excitations. This dual nature has been observed in previous tunneling experiments[30] and is consistent with our observations of sub-gap transport all the way to zero bias.

We shall now discuss the origin of the conductance features that appear in transport measurements. Our starting point is the single-level QD Hamiltonian

$$H_{QD} = \sum_{\sigma=\{\uparrow,\downarrow\}} \varepsilon_\sigma n_\sigma + U n_\uparrow n_\downarrow, \qquad (1)$$

where $n_\sigma = d_\sigma^+ d_\sigma$ is the electron number operator, $d_\sigma^+ (d_\sigma)$ creates (annihilate) an electron with spin $\sigma$ on the QD, $\varepsilon_\sigma$ is the single-electron energy on the QD (which is tuned by $V_{sg}$ and $B$ field), and $U$ is the electron interaction parameter that can be both positive (repulsive) and negative (attractive). As we have described in the introduction, in the zero magnetic field ($\varepsilon_\uparrow = \varepsilon_\downarrow$) the parity of the QD ground and first excited state depends on the sign of interactions. Specifically, for the case of attractive interactions ($U<0$) the QD ground state has even parity as does the first excited state and the odd parity states lie at higher energies [see Fig. 1(a)].

How does the unusual level structure in the presence of attractive interactions on the QD reflect on transport through the QD? We begin by considering the case in which both the superconducting and the normal leads are weakly coupled to the QD. In this case the electrons move by a series of resonant pair tunneling processes: the electron pair tunnels from the source lead to the QD and then to the drain lead. In order for the resonant tunneling processes to take place the two-electron excitation on the QD must be resonant with an occupied two-electron state in the source lead and an empty two-electron state in the drain lead. The two-electron spectral function in a superconductor has a $4\Delta$ gap, as compared to the one-electron spectral function that has a $2\Delta$ gap. Taking into account this gap we find the conductance maps (see Fig. 3). We observe that in order to connect the two diamonds with a straight line, as seen in the experiment, we must have one lead gapless, resulting in a $4\Delta/e$ gap as shown in Fig. 2(a).

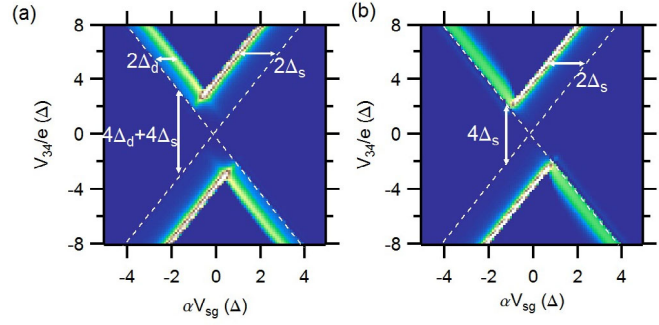

FIG. 3, Simulation of pair conductance diamonds on varying gapped excitations in the leads. (a) When both source and drain leads only have gapped excitations, the diamonds offset away from the gapless excitations indicated by the dashed lines. An insulating gap of $4(\Delta_s+\Delta_d)/e$ appears between the tips of diamonds, where $\Delta_s$ and $\Delta_d$ are the pairing gaps of source and drain leads. (b) When the drain lead has gapless excitations, one side of the diamonds stay connected by a straight line.

As the coupling between the QD and the superconducting lead becomes stronger, the QD begins to coherently exchange electrons with the superconductor. We describe these processes by supplementing $H_{QD}$ with $H_{SC}$ that describes the conventional gapped Bolgoliubov excitations in the superconducting lead, and $H_T$ that describes the electron tunneling between the superconducting lead and the QD

$$H = H_{SC} + H_{QD} + H_T, \tag{2}$$

$$H_{SC} = \sum_{k\sigma} \xi_k c^+_{k\sigma} c_{k\sigma} + \Delta \sum_k (c^+_{k\uparrow} c^+_{-k\downarrow} + c_{-k\downarrow} c_{k\uparrow}), \tag{3}$$

$$H_T = \sum_{k\sigma} t c^+_{k\sigma} d_\sigma + h.c., \tag{4}$$

Where $c^+_{k\sigma}$ and $c_{k\sigma}$ are the electron creation and annihilation operators in the superconducting lead, $\xi_k$ is the electron energy in the absence of the pairing gap $\Delta$, and $t$ is the tunneling coefficient.

The experimentally-observed sub-gap features can be readily seen in the one- and two-electron density of states (DOS) computed within our model (see Appendix II and III for details).

For the case of strong attractive interactions ($U < -\Delta$), only the two-electron spectral function has sub-gap features. These "X"-shaped features originate in pair bound states on the QD and have particle-hole symmetry [see Fig. 4(a)]. On the other hand, for the case of strongly repulsive interactions ($U > \Delta$) only the one-electron spectral function has sub-gap features, and these originate in the ABS [see Fig. 4(b)]. The qualitative appearance of these sub-gap features is not sensitive to details such as the tunneling strength $t$ or the exact value of the interaction strength $U$. By comparing the sub-gap spectral function features with the experimental transport data we can identify two regimes in the transport data: the pair bound state regime and the ABS regime. We therefore identify the experimentally-observed transition in the character of transport with the change in the sign of electron-electron interactions on the QD.

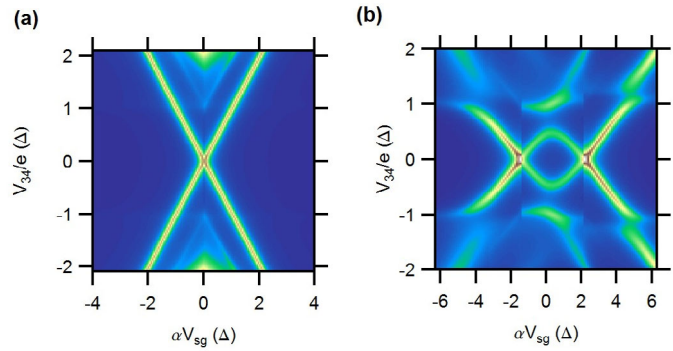

FIG. 4, Theoretical calculation of DOS spectra in a single level QD in the presence of (a) attractive ($U$=-4Δ) and (b) repulsive ($U$=2Δ) electron-electron interaction. For the case (a) of strong attractive interactions, the two-electron "X"-shaped resonances are dominant, whereas for case (b) of strong repulsion, the dominant sub-gap "loop" features are one-electron resonances with Andreev bound states.

## IV. MECHANISMS FOR DENSITY-TUNED INTERACTIONS – LIFSHITZ TRANSITION AND OTHER ALTERNATIVES

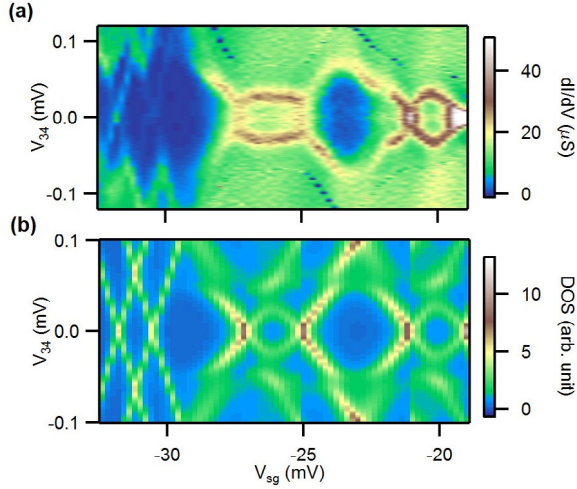

FIG. 5, Comparison between data and calculation. (a) Magnified data plot in -33 mV<$V_{sg}$<-19 mV. (b) Calculation of the DOS on the QD in the same $V_{sg}$ range. The QD is restricted to 4 levels, with negative (positive) interaction for the bottom (upper) 2 levels in band 1 (2).

To model the experimentally observed transition from attractive to repulsive interactions, we extend the QD to 4 levels with the lower 2 levels of attractive character and the upper 2 levels of repulsive character. The corresponding one- and two-electron spectral functions [see Fig. 5(b)] show two distinct regimes: "X"-shaped two-electron features at low electron densities on the QD, and loop-shaped features at high electron densities. The simple 4-level QD calculation agrees with the experimental data quite well [see Fig. 5(a)].

While electron-electron interactions are generally tuned by the electron density, it is important to consider why the observed transition from attractive to repulsive interactions such an abrupt function of the electron density. We suspect that the underlying mechanism is connected to the Lifshitz transition at the LAO/STO interface. The 2DEG at interface is formed from the three titanium $t_{2g}$ d electron bands. Interfacial confinement effects split these d electron bands into a lower $d_{xy}$ band and higher $d_{xz}/d_{yz}$ bands[31]. We conjecture that the $d_{xy}$ electrons have attractive character while the $d_{xz}/d_{yz}$ electrons have repulsive character. At low electron densities only the $d_{xy}$ levels are available and hence the interactions on the QD are attractive. At a critical electron density, marked by the Lifshitz transition point, the higher $d_{xz}/d_{yz}$ bands become available and the interactions on the QD become repulsive. This interpretation that the lower $d_{xy}$ band is the cradle of attractive interactions is consistent with the measurement at the 2D LAO/STO interface, which shows that the optimal doping for superconductivity happens at the Liftshitz transition[31].

We now consider alternative explanations for the abrupt change in the character of transport. Abruptly increasing the tunneling matrix element $t$ (e.g. by gating the barrier between the QD and the superconducting lead) may seem like a viable candidate for affecting the ground state parity[27], but an increase in $t$ (with increasing $V_{sg}$) neither favors an odd parity ground state nor does it bring down the single-electron states into the gap, which conflicts with the observation here. A more workable possibility is to abruptly introduce a large Zeeman field, in the presence of attractive interactions, to break the electron pairs on the QD and thus drive a transition from the two-electron to the one-electron transport regime. However, the only possible origin of such a Zeeman field is the exchange interaction between electron spins on the QD and a magnetic impurity spin in a charge trap. Loading an electron into the charge trap has a large impact on the transport characteristics[24, 32, 33], either giving rise to a sudden "sawtooth like" diamond if the trap is in parallel with the QD[33, 34] or causing a large insulating gap independent of the opening and closing of the pairing gap inside the diamonds if the trap is in series with the QD. Because these trap signatures are not observed here, it is highly unlikely that the transition could be attributed the transition to the presence of impurity spin.

## V. SIGNATURES OF PREFORMED PAIRS

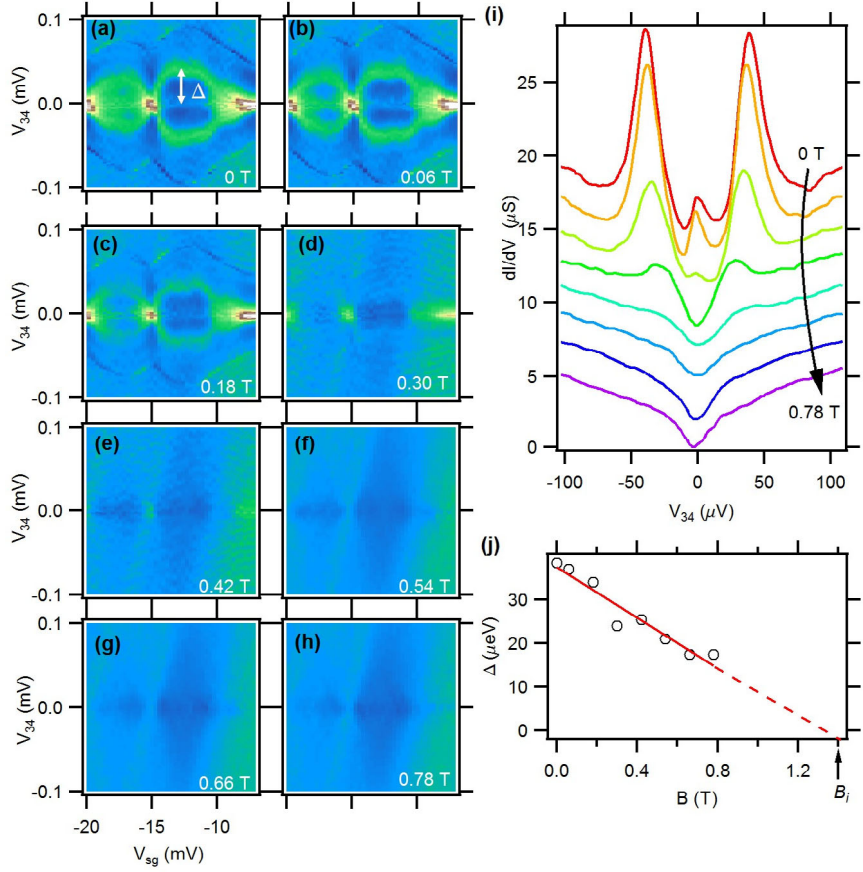

FIG. 6, Low-field dependence of ABS. (a)-(h), ABS loops at $B$=0 T, 0.06 T and 0.18 T to 0.78 T in step of 0.12 T. (i), Average vertical line cuts (averaged in -14 mV<$V_{sg}$<-11 mV). Curves are shifted for clarity. (j), Extracted pairing gap size as function of $B$.

So far we have discussed our observations of ABS at the strongly correlated LAO/STO interface. In other strongly correlated systems like high-$T_c$ cuprates, ABS is predicted to exist in the pseudogap regime[35]. We now explore the correlation between ABS and pre-formed pairs in LAO/STO by studying the low-magnetic-field dependence of ABS loops. As shown in Fig. 6(a)-(h), the amplitude and width ($2\Delta$ in $V_{34}$ direction) of the ABS loops shrink with increasing magnetic field. This evolution is more clearly visible by examining the average line-cuts in the range $-15\text{ mV} < V_{sg} < -10\text{ mV}$ [see Fig. 6(i)]. The ABS peaks are completely suppressed above $\mu_0 H_{c2} = 0.3\text{ T}$. The remaining dip at zero bias is an indication of the pairing gap at higher fields.

At $B < \mu_0 H_{c2}$, additional ZBP features appear inside the loops and carry supercurrent at $V_{sg} = -20$ mV, -15 mV, and -6 mV where the QD levels align with the source and drain chemical potentials. These features are a consequence of coherent pair tunneling across the QD and are not present in every device. The extracted pairing energy (for the lead) decreases linearly with increasing field, with a zero-energy field intercept $B_i = 1.3$ T which is inconsistent with $B_p$ in the lower $V_{sg}$ regime.

## VI. CONCLUSIONS

The sign of the electron-electron interaction at the LAO/STO interface has a profound influence on the electron transport in SSET devices. The attractive interaction in the low $V_{sg}$ regime results in electrons tunneling in pairs even at conditions where superconductivity is suppressed. Meanwhile, the emergence of single-particle ABS loops in the high $V_{sg}$ regime is characteristic of repulsive electron-electron interactions. This abrupt sign change of electron-electron interactions, tuned by a single parameter $V_{sg}$, is postulated to be driven by the discontinuity of band structure at the Lifshitz transition.

The novel reconfigurability of nanostructures at the LAO/STO interface has already provided an essential capability for engineering quantum states. The successful observation of tunable electron-electron interactions adds another key piece to the "Hubbard toolbox" for solid state quantum simulation[36].

ACKNOWLEDGEMENTS

This work is supported by the Air Force Office of Scientific Research under grant No. FA9550-10-1-0524 (J.L.) and FA9550-12-1-0057 (J.L., A.D., A.B.T.), the National Science Foundation under grant No. DMR-1104191 (J.L.), the Office of Naval Research under grant No. N00014-15-1-2847 (J.L.). Work at the University of Wisconsin was financially supported by the DOE Office of Basic Energy Sciences under award No. DE-FG02-06ER46327.

**APPENDIX I: TRANSPORT IN THE WELL-DEFINED CONDUCTANCE DIAMONDS REGIME**

In the well-defined conductance diamonds regime, the strong electron-electron attraction dominates the spectrum of the QD. Therefore, we treat the electrons on the QD as being tightly bound into pairs, and low energy excitations of the QD correspond to adding or removing an electron pair from the QD. The effective Hamiltonian for the QD becomes

$$H_{QD} = (C_{sg}V_{sg} - 2ne)^2 / C_{\Sigma}, \qquad (5)$$

where $C_{sg}$ and $C_{\Sigma}$ are the effective gate capacitance and total capacitance for adding electron pairs, and $n$ is the number of pairs on QD. We model the transport through the QD using a master equation that describes the hopping of electron pairs between the leads and the QD. To connect the QD to the leads we need the two-electron spectral functions $A_1^{(2)}(\omega)$ and $A_2^{(2)}(\omega)$ in the two superconducting leads along with the pair distribution functions. We can split the spectral function in the leads into three contributions[37]:

(1) a peak at $\omega = 0$ corresponding to the pair condensate (this peak is expected to be significantly broadened for 1D superconductors, like our leads);

(2) a finite spectral weight for $\omega < 2\Delta$ corresponding to bound pairs at finite momentum (i.e. the phase and amplitude modes);

(3) a large spectral weight at $\omega \geq 2\Delta$ corresponding to pairs of free propagating particles (either hole-like or electron-like).

Instead of computing the spectral function and the pair distribution function from first principles, we use a phenomenological model. To account for the fact that the pairs are made of electrons, we use the Fermi-distribution function $n_F$ to model the pair distribution function. We model the spectral function using the expression

$$A_j^{(2)}(\omega) = \text{Re}\left(\frac{1}{\sqrt{\omega^2 - (2\Delta_j)^2 + i\gamma_j^2}}\right), \tag{6}$$

which has peaks at $\omega = \pm 2\Delta$ associated with type (3) excitation and a finite weight at $0 \leq \omega < 2\Delta$ associated with type (1) and (2) excitations.

Consider the $V_{sg}$ range near the tip of one of the conductance diamonds where the QD level with $n+1$ pairs becomes degenerate with the QD level with $n$ pairs. The populations with $n$ and $n+1$ pairs on the QD follow

$$\dot{c}_n = -c_n \sum_{j=1,2} A_j(\mu_j - \varepsilon) n_F(\mu_j - \varepsilon) + c_{n+1} \sum_{j=1,2} A_j(\mu_j - \varepsilon)(1 - n_F(\mu_j - \varepsilon)), \tag{7}$$

$$\dot{c}_{n+1} = c_n \sum_{j=1,2} A_j(\mu_j - \varepsilon) n_F(\mu_j - \varepsilon) - c_{n+1} \sum_{j=1,2} A_j(\mu_j - \varepsilon)(1 - n_F(\mu_j - \varepsilon)), \tag{8}$$

where $\mu_1 = eV_{34}/2$ and $\mu_2 = -eV_{34}/2$ are the chemical potentials in the two leads and $\varepsilon = \alpha(V_{sg} - V_{sg0})$ converts $V_{sg}$ to energy with the lever arm $\alpha$ and $V_{sg0}$ is the degeneracy point between states with $n$ and $n+1$ pairs on the. The corresponding current is

$$I(\mu_1, \mu_2, \varepsilon) = \frac{A_1^{(2)}(\mu_1 - \varepsilon) A_2^{(2)}(\mu_2 - \varepsilon)[n_F(\mu_1 - \varepsilon) - n_F(\mu_2 - \varepsilon)]}{A_1^{(2)}(\mu_1 - \varepsilon) + A_2^{(2)}(\mu_2 - \varepsilon)}. \tag{9}$$

d$I$/d$V$ obtained from this formula is plotted in Fig. 3.

# APPENDIX II: SPECTRAL FUNCTIONS

Following Eqs. (2)-(4), we work in the Bogoliubov quasi-particle representation with $\xi_k = \hbar^2 k^2/(2m^*) - E_F$, where $E_F$ is the Fermi energy and $m^*$ is the effective mass of the electron. The creation and annihilation operators can be written as

$$c_{k\uparrow} = u_k \gamma_{k\uparrow} + \upsilon_k \gamma_{k\downarrow}^+, \tag{10}$$

$$c_{-k\downarrow} = u_k \gamma_{k\downarrow} - \upsilon_k \gamma_{k\uparrow}^+, \tag{11}$$

where $u_k = \sqrt{\frac{1}{2}(1+\frac{\xi_k}{E_k})}$ and $\upsilon_k = \sqrt{\frac{1}{2}(1-\frac{\xi_k}{E_k})}$. This brings $H_{SC}$ to diagonal form

$$H_{SC} = \sum_{k\sigma} E_k \gamma_{k\sigma}^+ \gamma_{k\sigma}, \tag{12}$$

where $E_k = \sqrt{\Delta^2 + \xi_k^2}$. Then we can write $H_T$ as

$$H_T = \sum_{kj\sigma}[t_j (u_k \gamma_{k\sigma}^+ + \sigma \upsilon_k \gamma_{k\bar{\sigma}}) d_{j\sigma} + h.c.], \tag{13}$$

where the tunneling coefficients $t_j$ only depend on the quantum dot's energy level $j$. We then numerically reconstruct the QD's DOS by computing the one- and two-electron spectral functions, which are given by

$$A_{j,\sigma}^{(1)}(V) = \sum_n (|\langle \psi_n | d_{j\sigma} | \psi_g \rangle|^2 \, \delta(E_n - E_g - eV) + |\langle \psi_n | d_{j\sigma}^+ | \psi_g \rangle|^2 \, \delta(E_n - E_g - eV), \tag{14}$$

$$A_{i,j}^{(2)}(V) = \sum_n (|\langle \psi_n | d_{i\uparrow} d_{j\downarrow} | \psi_g \rangle|^2 \, \delta(E_n - E_g - eV) + |\langle \psi_n | d_{i\uparrow}^+ d_{j\downarrow}^+ | \psi_g \rangle|^2 \, \delta(E_n - E_g - eV), \tag{15}$$

where $|\psi_g\rangle$ represents the ground state of the composite S-QD system and $\{|\psi_n\rangle\}$ the manifold of excited states, with $E_g$ and $\{E_n\}$ being their respective energies. The QD's DOS is then given by

$$N_{QD}(V) = \sum_{j,\sigma} A_{j,\sigma}^{(1)}(V) + \sum_{i,j} A_{i,j}^{(2)}(V) \tag{16}$$

In the calculations of this work, we account for broadening effects by replacing the delta functions in Eqs. (9) and (10) for (unity normalized) Lorentzians with width $\Gamma$ of the form

$$\delta(E_e - E_g - eV) \to \frac{\Gamma/(2\pi)}{(E_e - E_g - eV)^2 + (\Gamma/2)^2}. \qquad (17)$$

## APPENDIX III: NUMERICAL CALCULATION OF THE DOS

In tunnel experiments, one can typically express the tunneling current in terms of the spectral functions. In particular, if the DOS of the tunneling probe can be assumed to be approximately constant, one can show that to lowest order in the tunneling [38]

$$\frac{dI}{dV} \propto \sum_{j,\sigma} A^{(1)}_{j,\sigma}(-eV) \qquad (18)$$

which allows for a direct mapping between the one-electron DOS of the device and the measured d$I$/d$V$.

We numerically reconstruct the QD's DOS by diagonalizing the model Hamiltonian as a function of chemical potential $\mu(V_{sg})$ to compute the one- and two-electron spectral functions, as instructed by Eq. (16). We first consider the superconductor's quasiparticle modes in the continuum limit, so that

$$H_{SC} = \sum_\sigma \int_\Delta^\infty dE\, \gamma_\sigma^\dagger(E) E \gamma_\sigma(E), \qquad (19)$$

$$H_T = \sum_{j,\sigma} t_j \int_\Delta^\infty dE\, g(E)\left(u(E)\gamma_\sigma^\dagger(E) + \sigma v(E)\gamma_{\bar\sigma}(E)\right) d_{j,\sigma} + h.c., \qquad (20)$$

where $\gamma_\sigma(E) = g(E)\gamma_{k\sigma}$ and

$$g(E) = \sqrt{\frac{L}{2\pi}\frac{dk}{dE}} = \left(\frac{L}{2\pi}\frac{\sqrt{m}}{\sqrt{2}\hbar}\frac{E}{(E^2-\Delta^2)^{3/4}}\right)^{1/2} \qquad (21)$$

with $L$ being the length of the superconducting wire. We then discretize the energy integrals and

the energy-dependent quasi-particle operators into $M$ effective modes according to

$$\int_{E_i}^{E_{i+1}} dE f(E) \cong \varepsilon f(E_{i+1/2}), \tag{22}$$

$$\gamma_\sigma(E_{i+1/2}) = \gamma_{i\sigma}/\sqrt{\varepsilon}, \tag{23}$$

where

$$\varepsilon = \frac{E_c - \Delta}{M} \tag{24}$$

is the energy spacing between two consecutive quasiparticle levels, defined in terms of an energy cutoff $E_{\text{cut}}$. Putting these results together gives the final form of the discretized superconductor and tunneling Hamiltonians

$$H_{SC} = \sum_\sigma \sum_{i=1}^{M} E_{i+1/2} \gamma_{i\sigma}^\dagger \gamma_{i\sigma}, \tag{25}$$

$$H_T = \sum_{j,\sigma} \sum_{i=1}^{M} \tau_{ij} \left( u(E_{i+1/2}) \gamma_{i\sigma}^\dagger + \sigma v(E_{i+1/2}) \gamma_{i\bar\sigma} \right) d_{j,\sigma} + h.c., \tag{26}$$

where

$$\tau_{ij} = t_j \sqrt{\varepsilon} g(E_{i+1/2}) = \tilde{t}_j \left( \frac{\varepsilon E_{i+1/2}/\Delta^2}{\left(E_{i+1/2}^2/\Delta^2 - 1\right)^{3/4}} \right)^{1/2}, \tag{27}$$

with

$$\tilde{t}_j = t_j \left( \frac{L}{2\pi} \frac{\sqrt{m\Delta}}{\sqrt{2\hbar}} \right)^{1/2} \tag{28}$$

which we treat as a free parameter. Other free parameters include the QD's energies $\varepsilon_{j\sigma}$ and the interaction coefficients $U_{ij}$, which we adjust in order to reproduce the subgap features in the observed d$I$/d$V$ characteristics shown in Fig. 4(a). We use the experimental estimate of $\Delta = 48$ $\mu$eV for the superconducting gap (at $V_{sg} = -40$ mV) and assume a linear relationship between $V_{sg}$ and $\mu$, phenomenologically found to be approximately given by $\mu \cong eV_{sg}/20$. The calculated DOS is shown in Fig. 4(b). This simulation is for a 4-level QD, with two levels lying within each band, with electrons in band 1 being strongly attracting ($U_1 < 0$) and in band 2 repulsive ($U_2 > 0$). We also allow for interband interactions ($U_{12} \neq 0$). To make this calculation numerically tractable, we reduce the size of the Hilbert space of the SC to the one- and the two-quasiparticle sectors, with the latter being restricted to the subspace of two-quasiparticle states of opposite spins. In addition, we further reduce the size of the total Hamiltonian matrix by only considering the coupling between states whose overall energies lie within the energy window set by the energy cutoff $E_{cut} = 6\Delta$. The broadening of resonance lines is qualitatively captured by replacing the delta functions by Lorentzians in the spectral functions and by adjusting the width $\Gamma$.

## APPENDIX IV: RCSJ MODEL

At sufficiently high $V_{sg}$ values ($V_{sg} > -10$ mV), the two barriers become transparent and coherent Josephson transport becomes dominant. The $I$-$V$ curves can be well fitted by the extended resistively and capacitively shunted junction (RCSJ) model[28, 29]. We take into account the lead resistance $R_L$ (of wire sections from the barriers to lead 3 and 4) and shunt resistance $R_J$ of the QD [Fig. 7]. The $I$-$V$ curve takes the following form

$$I(V_{34}) = \{I_c \, \text{Im}[\frac{I_{1-i\eta}(I_c\hbar/2ek_BT)}{I_{-i\eta}(I_c\hbar/2ek_BT)}] + \frac{V_{34}}{R_J}\} \frac{R_J}{R_J + R_L} \qquad (29)$$

Where $\eta = \hbar V_{34}/2eRk_BT$, $k_B$ is the Boltzman constant and $I_\alpha(x)$ is the modified Bessel function of complex order $\alpha$. The extracted critical current $I_c = 2.8$ nA (at) is larger than the switch $V_{sg} = 0$ mV current $I_s = 2.8$ nA. Theoretically, the maximum of critical current $I_{cmax}$ has a simple relation with $\Delta$ in the strong-coupling regime, $I_{cmax} = 2\pi\Delta e/h$ by assuming equal coupling strength of two barriers, where $h$ is the Planck constant[39]. Taking $\Delta = 48$ $\mu$eV, the calculated $I_{cmax} = 11.7$ nA is about 4 times of the measured result. This is in fact in excellent agreement considering only a room temperature microwave (RF) filter is used in the experiment, as electromagnetic radiation is the major reason for this discrepancy.

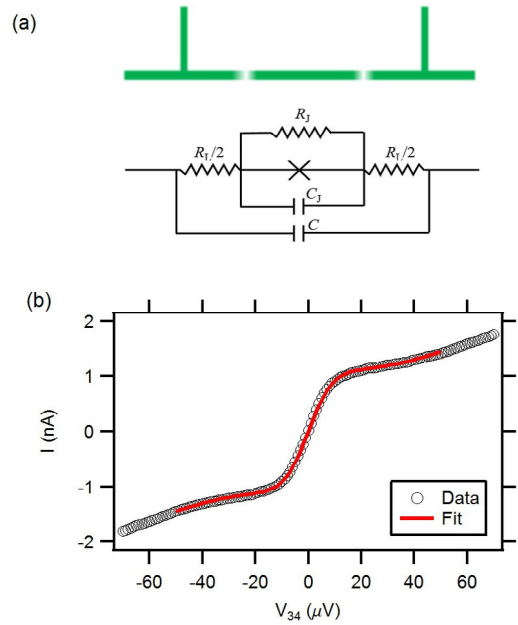

FIG. 7 RCSJ model fitting. (a) Schematic. (b) RCSJ fitting of I-V curve at $V_{sg}$=0 mV yielding $I_c$=2.8 nA, $R_J$=40.4 kΩ and $R_L$=5.0 kΩ.